\documentclass[letter]{./aa}  
\usepackage{natbib}
\usepackage{graphicx}
\begin{document}
   \title{Strong horizontal photospheric magnetic field \\
          in a surface dynamo simulation}
   \titlerunning{Strong horizontal photospheric magnetic field}
   \author{M. Sch{\"u}ssler \inst{1} 
           \and A. V{\"o}gler \inst{2}
          }
 
   \institute{Max-Planck-Institut f\"ur Sonnensystemforschung, 
              Max-Planck-Strasse 2, 37191 Katlenburg-Lindau, Germany
              \and 
              Sterrekundig Instituut, Utrecht University,
              Postbus 80 000, 3508 TA Utrecht, The Netherlands \\
              \email{A.Voegler@astro.uu.nl, schuessler@mps.mpg.de}
             }

   \date{\today}

  \abstract
  {Observations with the {{\it Hinode\/}} spectro-polarimeter have revealed
  strong horizontal internetwork magnetic fields in the quiet solar
  photosphere.}
  {We aim at interpreting the observations by means of results from
  numerical simulations. }
  {Radiative MHD simulations of dynamo action by near-surface
  convection are analyzed with respect to the relation between vertical
  and horizontal magnetic field components. }
  {The dynamo-generated fields show a clear dominance of the horizontal
  field in the height range where the spectral lines used for the
  observations are formed. The ratio between the averaged horizontal and
  vertical field components is consistent with the values derived from
  the observations. This behavior results from the intermittent nature
  of the dynamo field with polarity mixing on small scales in the
  surface layers. } 
  {Our results provide further evidence that local near-surface dynamo
   action contributes significantly to the solar internetwork fields.}
   \maketitle
%

\section{Introduction}

The ubiquitous existence of small-scale `internetwork' magnetic fields
of mixed polarity in the so-called quiet solar photosphere is strongly
indicated by various observational diagnostics \citep[e.g.,][and further
references therein]{Khomenko:etal:2003, Lites:Socas-Navarro:2004,
Trujillo:etal:2004}.  Recent high-resolution space-borne observations
with the spectropolarimeter of the Solar Optical Telescope aboard the
{\it Hinode\/} satellite have considerably strengthened the case for
internetwork fields and, furthermore, have revealed that the measured
internetwork flux is dominated by strongly inclined, almost horizontal
magnetic field \citep{Lites:etal:2007, Orozco-Suarez:etal:2007}.
Considerable amounts of highly time-dependent horizontal magnetic flux
have also been found in ground-based observations with lower spatial
resolution \citep{Harvey:etal:2007}.  The ubiquity of the small-scale,
mixed-polarity internetwork field suggests a local origin of at least a
significant part of the measured flux.  Recently, we have demonstrated
by means of radiative magneto-convection simulations that local dynamo
action by near-surface convective flows is a possible source for the
internetwork flux \citep{Voegler:Schuessler:2007}. Here we show that the
spatial structure of the dynamo-generated field provides a natural
explanation for the observed dominance of the horizontal field component
in the middle photosphere.


\section{Numerical model}

We use  the results of  dynamo run C  of \citet{Voegler:Schuessler:2007}
with $648\times 648 \times 140$ grid cells in a computational box with a
physical size  of $4.86\times 4.86$~Mm$^2$ in the  horizontal and 1.4~Mm
in the vertical direction, the latter ranging from about 900~km below to
500~km  above the  average level  of  continuum optical  depth unity  at
630~nm wavelength $(\tau_{630}=1)$. The simulation has been run with the
{\sl  MURaM}   code  \citep{Voegler:2003,  Voegler:etal:2005}.   With  a
magnetic  Reynolds  number  of  about  2600,  the  simulation  shows  an
exponential  growth of a  weak seed  field with  an $e$-folding  time of
about 10  minutes. The magnetic energy  saturates at about  2.5\% of the
kinetic  energy of  the convective  flows, the  maximum of  the spectral
energy distribution lying at horizontal  spatial scales of a few hundred
km,  at which  scales  the field  displays  a distinctly  mixed-polarity
character.

\section{Relation of horizontal and vertical field components}

\begin{figure*}
\centering
\resizebox{0.95\hsize}{!}{\includegraphics{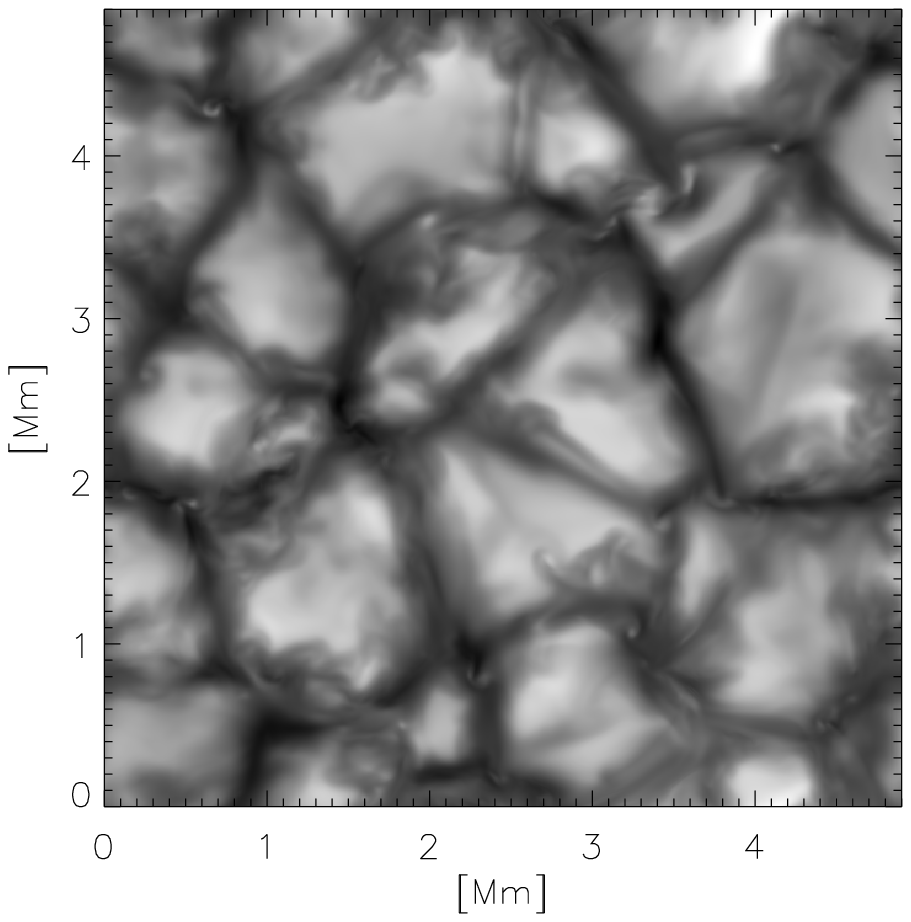}
          \includegraphics{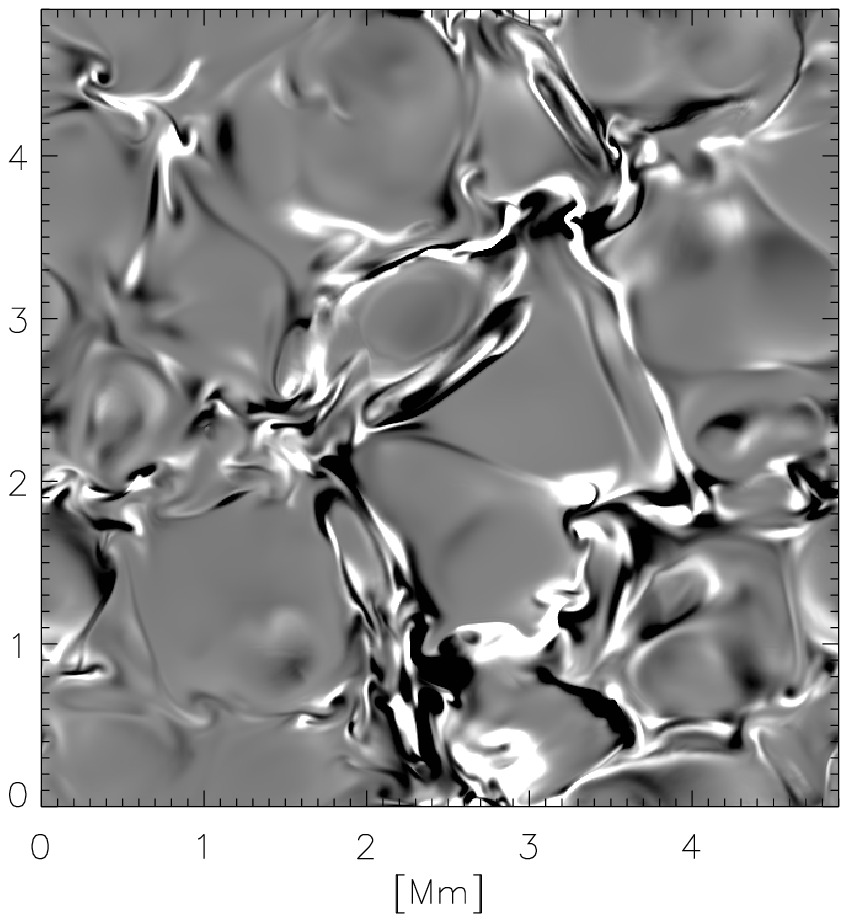}
          \includegraphics{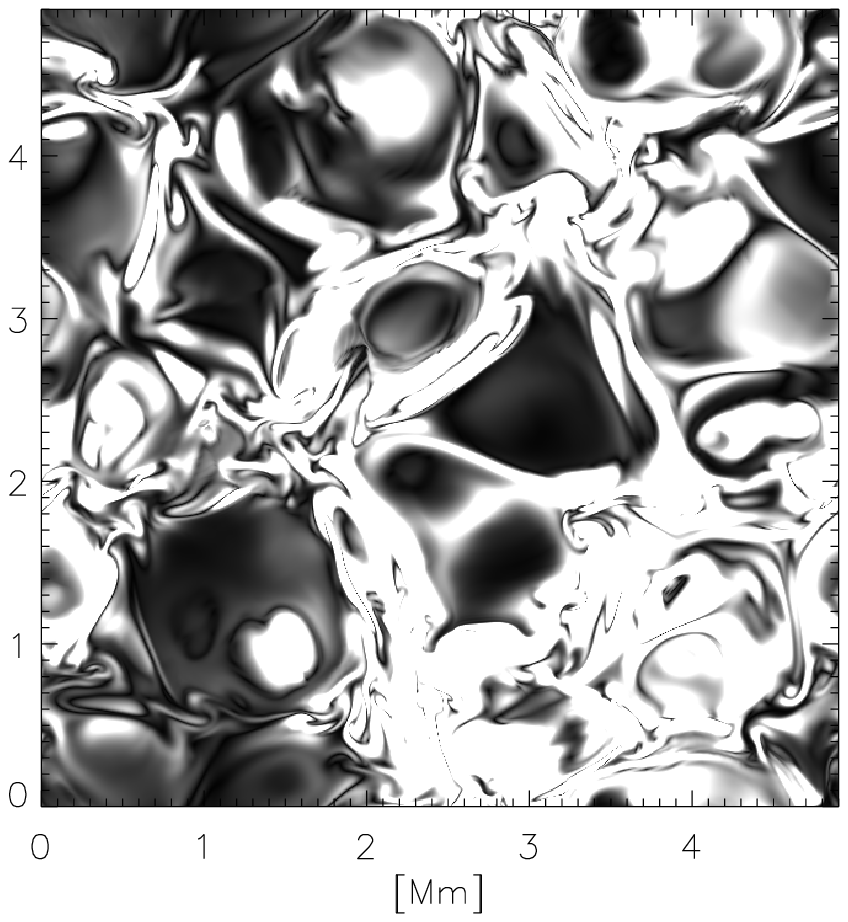}
                     }
\resizebox{0.95\hsize}{!}{\includegraphics{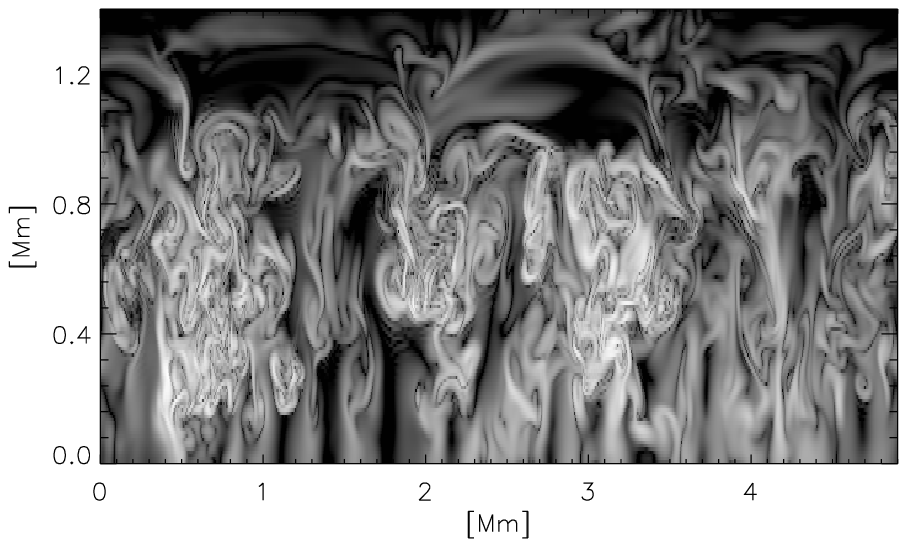}
          \hfill\includegraphics{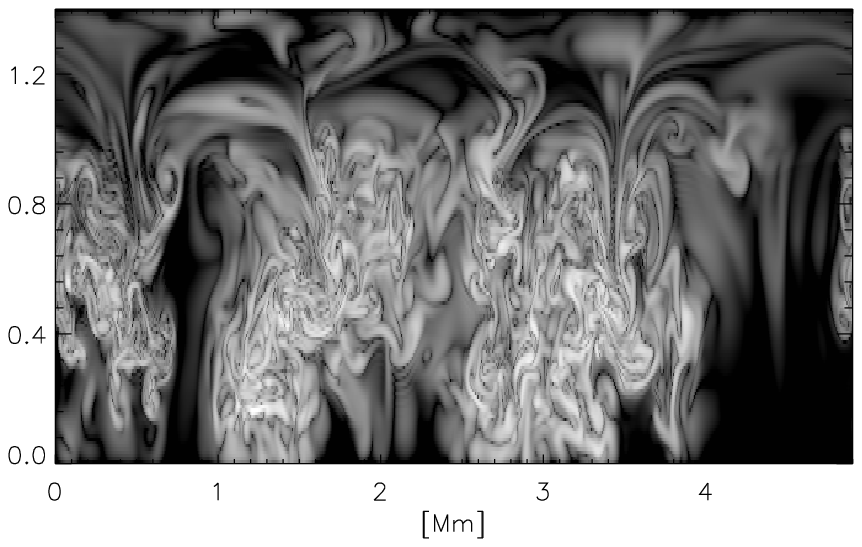}
                     }
\caption{Snapshot from the saturated phase of the dynamo run C
  \citep{Voegler:Schuessler:2007}. The panels in the upper row show
  the continuum intensity at 630~nm wavelength {\it (upper left)}, a
  gray-scale image (`magnetogram', saturated at $\pm 10\,$G, black and
  white colour indicating the two polarities) of the (signed) vertical
  field component on the surface $\tau_{630}=10^{-2}$, roughly
  corresponding to the formation height of the spectral line used for
  the observations {\it (upper middle)}, and the unsigned
  horizontal field strength {\it (upper right)}, black colour in this
  panel representing very weak field and white colour indicating the
  saturation level of $10\,$G. The two panels in the lower row show
  gray-scale images of the logarithm of the horizontal field strength
  (black colour indicating fields below $1\,$G) for vertical cuts
  through the simulation box along a horizontal line at $y=1.5\,$Mm
  (left) and $y=3\,$Mm (right), respectively, in the upper panels.  The
  level of $\tau_{630}=1$ is roughly at a height of $0.9\,$Mm, the level
  of $\tau_{630}=10^{-2}$ at $1.2\,$Mm.}
\label{fig:snapshot}
\end{figure*}

Figure~\ref{fig:snapshot} is based upon a snapshot from the saturated
phase of the dynamo run. The intensity image in the upper left panel
shows the granulation pattern, which is almost undisturbed by the
presence of the magnetic field. The distribution of the vertical field
component on the level surface $\tau_{630}=10^{-2}$ (upper middle panel)
reveals the mixed-polarity nature of the dynamo-generated magnetic
field, which preferentially resides in the intergranular downflow
lanes. Note that the distribution of vertical field at this level is
significantly smoother than at the height of optical depth unity
\citep[cf. Fig.~2 of][]{Voegler:Schuessler:2007}. This reflects the fact
that much of the small-scale flux at the lower level has already been
connected back by shallow loops. As a consequence, the unsigned vertical
field drops much more rapidly with height than the (unsigned) horizontal
field, so that at the level $\tau_{630}=10^{-2}$ the latter (shown in
the upper right panel) dominates in most places. The representation on
the two vertical cuts shown in the lower row of Fig.~\ref{fig:snapshot}
illustrates that the horizontal field in the photosphere (i.e., above a
height of about $0.9\,$Mm) has two components, namely, narrow loops near
the intergranular lanes and extended loops above granules.  The latter
have also been seen in the simulations of
\citet{Grossmann-Doerth:etal:1998} and \citet{Steiner:2007}; they are
presumably formed by reconnection events between loop `legs' with
opposite polarities and flux expulsion by the granular flows, larger
(stronger) granules pushing the horizontal field to higher levels in the
photosphere. As a consequence, a cut at a given level surface of
constant height (or optical depth) as shown in the upper right panel of
Fig.~\ref{fig:snapshot} misses part of the horizontal field above
granules and will be more dominated by the field around the
intergranular lanes.  \citet{Lites:etal:2007} report that the observed
horizontal field is spatially separated from the vertical field and
favors the edges of bright granules. A comparison with this finding
requires an analysis based upon synthetic Stokes profiles from the
simulation data, which will be presented in a later paper.

\begin{figure}
\centering
\resizebox{\hsize}{!}{\includegraphics{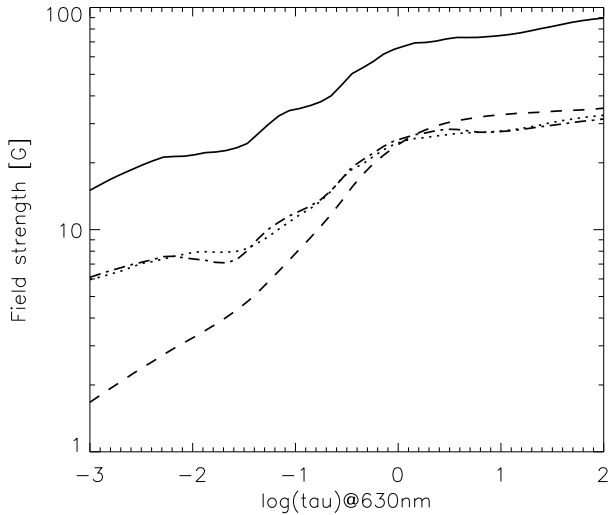}}
\caption{Profiles of magnetic field strength averaged over surfaces of
constant $\tau_{630}$: unsigned vertical field ($\langle|B_z|\rangle$,
dashed), unsigned horizontal field components
($\langle|B_{x,y}|\rangle$, dotted and dash-dotted, respectively), and
root-mean-square of the horizontal field strength ($\langle
B_x^2+B_y^2\rangle^{1/2}$, solid).
}
\label{fig:tauprof}
\end{figure}

A quantitative account of the relation between vertical and horizontal
field components is given in Fig.~\ref{fig:tauprof}. It shows field
strengths averaged over surfaces of constant optical depth as functions
of $\tau_{630}$. The dashed curve gives the average unsigned vertical
field while the dotted and dash-dotted curves represent the averages
of the two horizontal field components (unsigned).   

The averages of the three components indicate that the field is not far
from being statistically isotropic in the deep layers below
$\tau_{630}=1$. In contrast, the horizontal components of the magnetic
field become increasingly dominant in the photosphere above. 
The driving of the dynamo by small-scale turbulent shear flows
in and adjacent to the intergranular downflows is
mainly restricted to the regions below $\tau_{630}=1.$, whereas,
in the convectively stable photosphere above, these flows are much weaker and
the rate of work against the Lorentz force drops steeply with height.
Since inductive effects have smaller influence on the field in the
layers above $\tau_{630}=1$, the decay of the field with height is
mainly determined by its spatial structure at the surface (particularly
by the energy spectrum as a function of horizontal wavenumber).  This
results in a steep decline of the unsigned vertical field with height as
opposite polarities on small scales are connected by shallow loops with
typical length scales of a few hundred km, corresponding to the
horizontal scale for which the magnetic energy spectrum at
$\tau_{630}=1$ reaches its maximum.  It becomes plausible that this
configuration leads at the same time to a less steep decline of the
horizontal field, if one considers the simple example of an arcade-like
magnetic field with concentric semi-circular field lines:
for increasing height, more and more field
lines turn over horizontally, so that the horizontally averaged unsigned
vertical field strength decreases faster than the averaged horizontal
field; a simple calculation for this case shows that 
$\langle|B_{\rm hor}|\rangle$ strongly exceeds $\langle|B_{\rm vert}|\rangle$
at heights of the order of the horizontal scale (footpoint
separation at the surface) of the arcade. Thus, the dominance of 
horizontal fields in the photosphere is consistent with the assumption
of a simple loop topology with a preferred length scale.

The quantity that is actually relevant for a qualitative comparison with
the `apparent' horizontal field strength derived by
\citet{Lites:etal:2007} from measurements of the linear polarization
(transversal Zeeman effect, Stokes $Q$ and $U$) is the root-mean-square
of the horizontal magnetic field, i.e., $B_{\rm rms}\equiv\langle B_x^2
+ B_y^2\rangle^{1/2}$, since for not too strong fields Stokes $Q$ and
$U$ are proportional to the square of the horizontal field
strength. $B_{\rm rms}$ as a function of optical depth is shown as the
solid line in Fig.~\ref{fig:tauprof}. Owing to the inhomogeneity of the
horizontal field, this quantity is significantly larger than
$\langle|B_x|\rangle$ and $\langle|B_y|\rangle$, even if we multiply any
of them by a factor $\sqrt 2$ to take into account both horizontal field
components. From the ratio of the average horizontal field and the rms
field, we can estimate an average `fill fraction' as a measure of the
inhomogeneity of the horizontal field. We obtain a number between 0.25
and 0.3 in the range $-2 < \log\tau_{630} < -1$, which is the relevant
range of line formation of the FeI lines at 630.15~nm and 630.25~nm used
by the {\sl Hinode} spectro-polarimeter
\citep[e.g.,][]{Orozco:etal:2007b}. These values are consistent with the
estimate of the fill fraction obtained by \citet{Lites:etal:2007} and
\citet{Orozco-Suarez:etal:2007} by means of an inversion method.

\begin{figure}
\centering
\resizebox{\hsize}{!}{\includegraphics{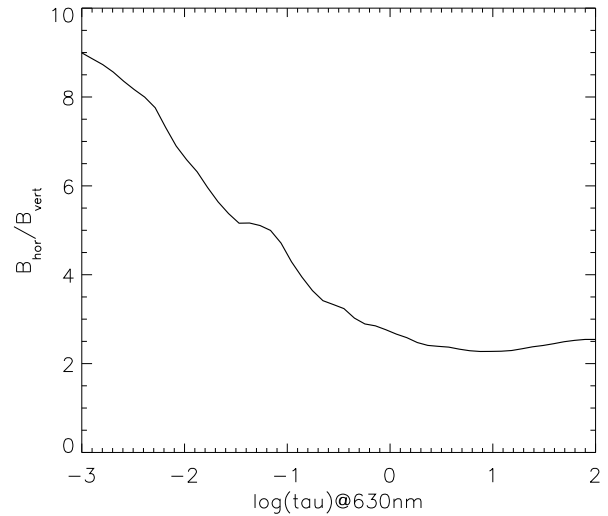}}
\caption{Ratio of the root-mean-square of the horizontal field component
  (solid line in Fig.~\ref{fig:tauprof}) to the averaged unsigned
  vertical field (dashed line in Fig.~\ref{fig:tauprof}). In the optical
  depth interval $-2 < \log(\tau_{630}) < -1$ (relevant for the
  formation of the FeI lines used in the {\it Hinode\/} SP), the ratio is
  roughly in the range 4--6.}
\label{fig:bratio}
\end{figure}

Let us now consider the ratio between the average vertical field and
$B_{\rm rms}$ as shown in Fig.~\ref{fig:bratio}. The ratio increases
strongly with height, so that in the optical depth interval $-2 <
\log\tau_{630} < -1$ values between 4 and 6 are reached. This is
consistent with the ratio of the horizontally averaged vertical
(longitudinal) and horizontal (transversal) apparent fields found by
\citet{Lites:etal:2007}: $B^L_{\rm app}/B^T_{\rm app} = 55\,{\rm
G}/11\,{\rm G} = 5$.  However, as pointed out by these authors, the
relation between $B^T_{\rm app}$ and the actual horizontal field
strength is far from trivial. Furthermore, effects of line-of-sight
integration and spatial smearing by the instrument complicate the
relationship between the average fields in the simulation and the field
strengths derived from the observed Stokes profiles.  A direct
quantitative comparison with the observations would have to proceed by
means of calculating synthetic Stokes profiles, taking into account the
point-spread function of the instruments. This is beyond the scope of
this {\sl Letter}.

\begin{figure}
\centering
\resizebox{\hsize}{!}{\includegraphics{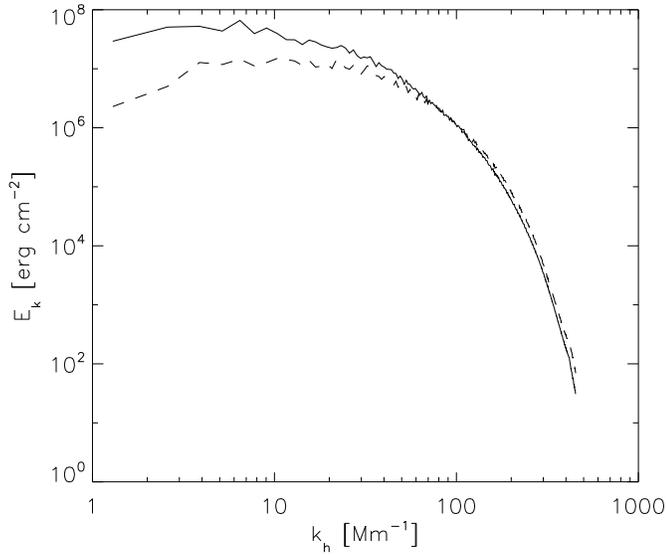}}
\caption{Magnetic energy spectra as a function of horizontal wave
  number, $k$, for the height range roughly corresponding to $-2 <
  \log\tau_{630} < -1$. The dashed curve shows the energy spectrum based
  on the vertical field component, while the solid curve gives the
  arithmetic mean of the spectra for the two horizontal field
  components. For low wavenumbers, the energy in the horizontal field
  clearly dominates.}
\label{fig:spectrum}
\end{figure}

\citet{Lites:etal:2007} have suggested that one possibility contributing
to the imbalance of the average vertical and horizontal fields could be
a significantly larger horizontal scale of the horizontal field as
compared to the vertical field. In fact, this is what we clearly find in
our dynamo simulation. Fig.~\ref{fig:spectrum} shows spectral magnetic
energy as a function of horizontal wave number. The dashed curve gives
the energy distribution for the vertical field, while the solid curve
represents the spectral energy in the horizontal field (mean of the
spectra for the two horizontal field components). For this plot we have
considered fields in the height range roughly corresponding to the
optical depth interval $-2 < \log\tau_{630} < -1$, which is relevant for
the formation of the iron lines used for the observations.  The curves
show that the field components are in equipartition at small scales
(large wave numbers), but that the horizontal field clearly dominates at
wave numbers below roughly 10 Mm$^{-1}$, corresponding to horizontal
scales larger than about 600~km.

\section{Discussion}
    
A direct comparison of the simulation results with the observations is
not possible because the individual values for the averaged fields in
the simulation are both (by about a factor three) smaller in the
relevant height range than the values for the `apparent' derived from
the observation, the discrepancy probably becoming even more severe when
the actual spatial resolution of the observations is taken into account.
This is not particularly surprising since the magnetic Reynolds number
of the simulation is still orders of magnitude smaller than the actual
value in the corresponding solar layers, so that the saturation level of
our simulation is probably considerably below the level to be expected
for the real Sun. In fact, a preliminary simulation with a roughly
doubled Reynolds number of about 5000 shows an increase of the magnetic
energy by a factor of about 1.7 with respect to the case shown
here. Interestingly, it turns out that the optical depth profiles of the
averaged field strengths in this case can be very well approximated by
just multiplying the curves shown in Fig.~\ref{fig:tauprof} by
$\sqrt{1.7}$, meaning that the main difference between the simulations
reduces to a simple scale factor for the field strength. Together with
the fact that the observed internetwork fields are predominantly weak
and their energy is significantly smaller than the kinetic energy of the
convective motions, this suggests that the general nature of the
dynamo-generated field may not be significantly different from the case
shown here, apart from a higher overall amplitude. In particular, we
expect that the ratio of the average horizontal and vertical field is
not strongly affected by the amplitude of the dynamo-generated field. Of
course, these assertions need to be demonstrated by further simulations
with higher Reynolds numbers.

The clear dominance of the horizontal field in the mid photosphere seems
to be a rather specific property of the strongly intermittent field
generated by near-surface turbulent dynamo action. Accordingly, the
dynamo simulation of \citet[][with a closed bottom boundary and a local
treatment of radiative transfer]{Abbett:2007} also exhibits strong
horizontal field in the photospheric layers. On the other hand, models
with an imposed net vertical flux \citep[e.g.,][]{Voegler:etal:2005} or
our recent simulations of the decay of a granulation-scale
mixed-polarity field (at magnetic Reynolds numbers below the threshold
of dynamo action) do not show this behavior; in these cases, the
intricate small-scale mixing of polarities that is characteristic for
the dynamo does not dominate the field structure. The rapid decay with
height of such a dynamo-generated field is also consistent with the
apparent lack of strong horizontal field in the chromosphere
\citep{Harvey:etal:2007}.

What are the alternatives to near-surface dynamo action?  `Shredding' of
pre-existing magnetic flux (remnants of bipolar magnetic regions) cannot
explain the large amount of observed horizontal flux since the turbulent
cascade does not lead to an accumulation of energy (and generation of a
spectral maximum) at small scales. On the other hand, such a behavior is
typical for turbulent dynamo action. Flux emergence from the deeper
convection zone in the form of granule-sized small bipoles would have to
proceed such a high rate in order to maintain the ubiquitous strong
horizontal fields that it probably would not have gone undetected in the
past \citep[see, however,][]{Centeno:etal:2007}. The
sporadic appearance of horizontal internetwork fields (HIFs) described
by \citet{Lites:etal:1996} and interpreted as small-scale flux emergence
events seems to be unsufficient to explain the ubiquitous horizontal
field now found with {\it Hinode}. On the other hand, flux recycling of
an overall background flux by granulation probably represents a
significant source of horizontal field in network and plage regions.
Emergence of extended horizontal field strands in granules as observed by
\cite{Ishikawa:etal:2007} is not seen in local dynamo simulations.

In the real Sun, probably all three sources, i.e., dynamo, shredded
fields, and small-scale flux emergence from deeper layers, contribute to
the internetwork flux in unknown amounts.  In any case, the strong
horizontal fields in the quiet photosphere inferred by the observations
indicate that the source of these fields at the solar surface is a
mixed-polarity field whose energy is mostly contained in those spatial
scales where the dynamo-generated flux resides.  Therefore, the
observational results obtained with the {\it Hinode\/ SP} together with
the analysis presented here provides strong indication that surface
dynamo action represents a significant source for the internetwork field
in the solar photosphere.


\bibliography{8998.bbl}

\end{document}